\documentclass[twocolumn]{aastex61}

\newcommand\aastex{AAS\TeX}

\shorttitle{\aastex\ CO diffusion and desorption kinetics in CO$_2$ ices}
\shortauthors{Cooke et al.}

\begin{document}

\title{CO diffusion and desorption kinetics in CO$_2$ ices}

\correspondingauthor{Ilsa R. Cooke}
\email{irc5zb@virginia.edu}

\author[0000-0002-0850-7426]{Ilsa R. Cooke}
\affil{Department of Chemistry \\
University of Virginia \\
McCormick Rd, Charlottesville,VA 22904, USA}
\affiliation{Harvard-Smithsonian Center for Astrophysics \\
60 Garden Street \\
Cambridge, MA 02138, USA}

\author[0000-0001-8798-1347]{Karin I. \"{O}berg}
\affiliation{Harvard-Smithsonian Center for Astrophysics \\
60 Garden Street \\
Cambridge, MA 02138, USA}

\author[0000-0001-8109-5256]{Edith C. Fayolle}
\affiliation{Harvard-Smithsonian Center for Astrophysics \\
	60 Garden Street \\
	Cambridge, MA 02138, USA}

\author{Zoe Peeler}
\affiliation{Harvard-Smithsonian Center for Astrophysics \\
	60 Garden Street \\
	Cambridge, MA 02138, USA}

\author[0000-0002-8716-0482]{Jennifer B. Bergner}
\affiliation{Harvard-Smithsonian Center for Astrophysics \\
	60 Garden Street \\
	Cambridge, MA 02138, USA}



\begin{abstract}
Diffusion of species in icy dust grain mantles is a fundamental process that shapes the chemistry of interstellar regions; yet measurements of diffusion in interstellar ice analogs are scarce. Here we present measurements of CO diffusion into CO$_2$ ice at low temperatures (T=11--23~K) using CO$_2$ longitudinal optical (LO) phonon modes to monitor the level of mixing of initially layered ices. We model the diffusion kinetics using Fick's second law and find the temperature dependent diffusion coefficients are well fit by an Arrhenius equation giving a diffusion barrier of 300 $\pm$ 40 K. The low barrier along with the diffusion kinetics through isotopically labeled layers suggest that CO diffuses through CO$_2$ along pore surfaces rather than through bulk diffusion. In complementary experiments, we measure the desorption energy of CO from CO$_2$ ices deposited at 11-50 K by temperature-programmed desorption (TPD) and find that the desorption barrier ranges from 1240 $\pm$ 90 K to 1410 $\pm$ 70 K depending on the CO$_2$ deposition temperature and resultant ice porosity. The measured CO-CO$_2$ desorption barriers demonstrate that CO binds equally well to CO$_2$ and H$_2$O ices when both are compact. The CO-CO$_2$ diffusion-desorption barrier ratio ranges from 0.21-0.24 dependent on the binding environment during diffusion. The diffusion-desorption ratio is consistent with the above hypothesis that the observed diffusion is a surface process and adds to previous experimental evidence on diffusion in water ice that suggests surface diffusion is important to the mobility of molecules within interstellar ices. 
 
\end{abstract}




\section{Introduction} \label{sec:intro}

The motion of atoms and molecules on and within icy grain mantles is a fundamental process that regulates the chemical evolution in astrophysical environments. Diffusion of these species within the bulk ice or along icy surfaces influences the rates of desorption, chemistry and ice reorganization. The interplay between diffusion and reaction of radical fragments within the ice is a critical factor to explain the existence and abundances of several complex organic molecules in star-forming regions \citep{Garrod2008,Garrod2013}.\\
\indent The diffusion of molecules in ice mantles is, however, poorly constrained. For most species and ice matrices, both the diffusion mechanism and the diffusion barrier are unknown. Astrochemical models therefore often adopt diffusion barriers that are fractions of the better constrained desorption barriers \citep{Tielens1982,Katz1999,Ruffle2000,Cuppen2009,Garrod2011,Chang2012}. Previous studies have explored diffusion-desorption barrier ratios between 0.3 and 0.8, and have demonstrated that the chemistry and ice composition is very sensitive to this parameter e.g. \citet{Garrod2011}; experimental constraints of diffusion and desorption for several major ice species are essential to better understand the temperature dependent evolution of ices. \\
\indent To obtain a complete understanding of diffusion in interstellar ices, data is required on diffusion in all common interstellar ice environments since molecular diffusion and desorption barriers are expected to depend strongly on the ice composition and morphology. Observations of ice absorption bands toward protostars indicate that the main ice constituents are H$_2$O, CO and CO$_2$. Furthermore, the ice mantles are typically divided into H$_2$O-rich and CO-rich phases, both of which are mixed with CO$_2$, as well as a pure CO$_2$ ice phase in some lines of sight \citep{DHendecourt1989,Boogert2004,Pontoppidan2008}. 
To understand the importance of diffusion in astrophysically relevant ices, experiments and models are required for all three ice phases. Diffusion of molecules through H$_2$O ices has been the subject of several recent studies \citep{Livingston2002,Mispelaer2013,Karssemeijer2014,Lauck2015}; however, diffusion in CO- and CO$_2$-rich ice environments has not been treated experimentally. Considering the differences between the ice matrices of CO- and CO$_2$-rich ices and the hydrogen-bonded, porous H$_2$O-rich ices, it is unclear whether the barriers and diffusion mechanisms in CO and CO$_2$ ices are similar to those found in the experiments with H$_2$O ices. \\
\indent Molecular diffusion in astrochemical ice analogs has been studied predominantly by two methods: firstly, by diffusion-desorption experiments in which the decreasing IR absorbance of the diffusing species is recorded over time \citep{Mispelaer2013,Karssemeijer2014}; and secondly, by spectroscopic techniques that exploit the fact that some IR bands are very sensitive to their molecular environment \citep{Lauck2015}. The latter effects have been shown to be strong when CO or CO$_2$ is mixed with hydrogen bonding molecules like water or methanol, producing blueshifts and broadening of the CO and CO$_2$ infrared modes \citep{sandford1988,Sandford1990,Ehrenfreund1999,Palumbo2000,Oberg2009}. However, the CO and CO$_2$ normal vibrational modes are not as sensitive to mixing with other, non-polar or weakly polar ice constituents, making diffusion measurements in these environments more challenging \citep{Ehrenfreund1997}. \\
\indent Recently, we have shown that CO$_2$ longitudinal optical (LO) phonons can be used to sensitively probe ice mixing characteristics including the amount of CO molecules that are mixed within CO$_2$ ices \citep{Cooke2016}. LO phonons arise in the CO$_2$ ice when the substrate is positioned at an oblique angle to the infrared beam. We found that the CO$_2$ LO phonons redshift linearly with the ice mixing fraction, suggesting that they may be utilized to study diffusion dynamics in CO$_2$ ices. \\
\indent Here, we present a study of CO diffusion into CO$_2$ ices by measuring changes in the CO$_2$ $\nu_3$ LO phonons. We also measure the desorption energy of CO from CO$_2$ ices and present the diffusion-desorption energy barrier ratio. 
Section \ref{sec:exptmntl} presents the experimental setup, procedures and spectral analyses used to study CO diffusion through CO$_2$ ices (\ref{subsec:Diffproc}) and desorption from CO$_2$ ices by temperature-programmed desorption (TPD) (\ref{subsec:TPDproc}).  Section \ref{sec:Diffresults} presents the results of the diffusion experiments and their dependencies on ice temperature and thickness as well as the diffusion modeling strategies. In section \ref{sec:TPDresults} we outline the results and analysis of the TPD experiments and extract the desorption barriers for CO from CO$_2$ ices. The results and their astrophysical implications are discussed in Section \ref{sec:Discuss}.

\section{Experimental Setup and Procedures}\label{sec:exptmntl} 
\subsection{Experimental Setup} \label{sec:setup}
\begin{figure*}[t!]
	\plotone{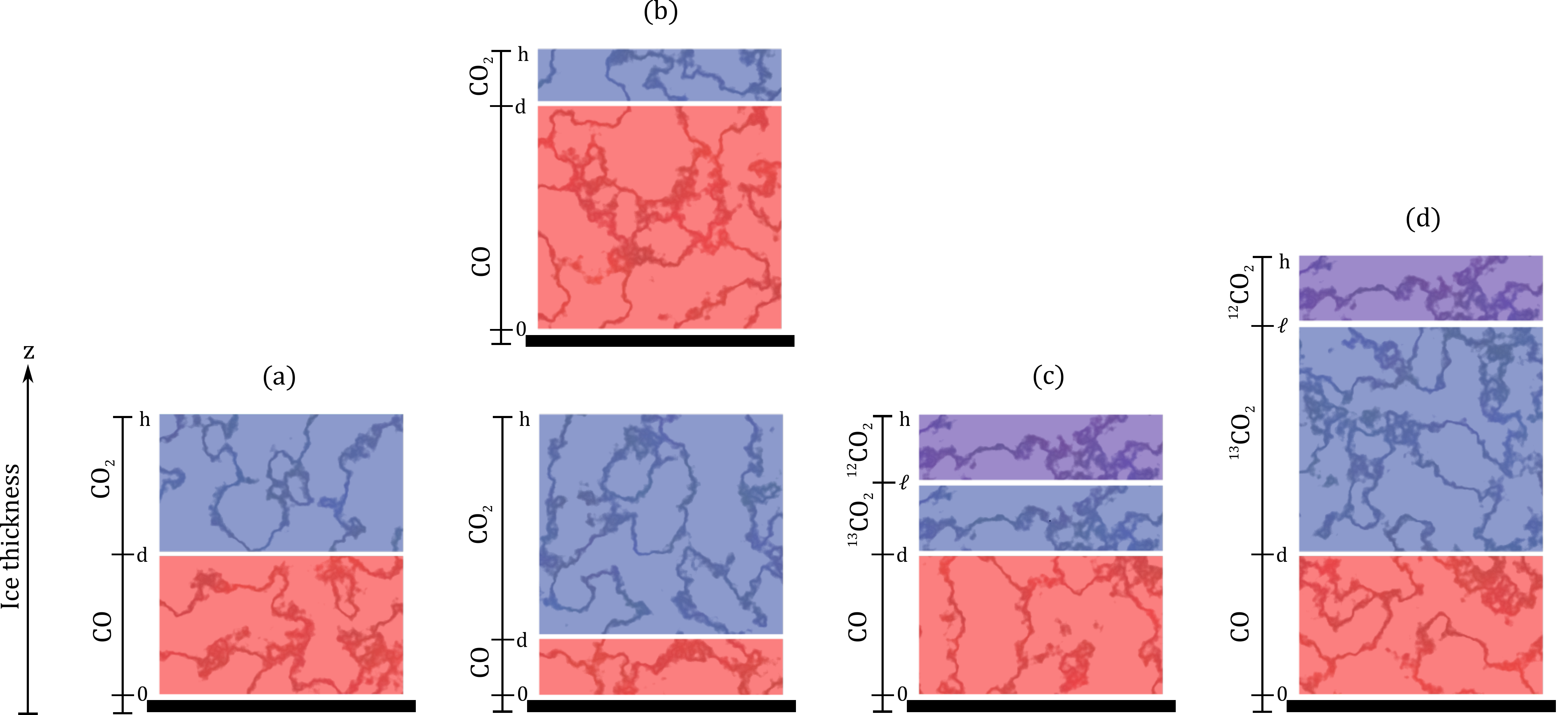}
	\caption{Schematic of ice configurations used during the diffusion experiments. The ices are displayed vertically along the z-axis, where \textit{d} represents the height of the CO-CO$_2$ interface and \textit{h} represents the height of the vacuum interface. \textit{(a)} In the fiducial experiment 30ML of CO$_2$ is layered on top of 30 ML CO and heated to 20 K. This configuration is repeated for diffusion temperatures in 1 K increments between 18-23 K. We also use this configuration but scale the two layers to 20 ML CO: 20 ML CO$_2$ to explore the ice thickness dependence. \textit{(b)} The ice thickness ratio was changed to 1:5 and 5:1 for the layered CO:CO$_2$ system. \textit{(c)} The 30 ML CO$_2$ ice was split into 15 ML $^{13}$CO$_2$ and 15 ML $^{12}$CO$_2$, $\ell$ represents the boundary between the two isotopologues. \textit{(d)} The thickness of the bulk isotopic layer was increased from 15 ML to 40 ML. }
	\label{fig:scheme}
\end{figure*}

The experiments were conducted using the setup described previously in \citet{Lauck2015}. Briefly, the setup consists of an ultrahigh vacuum chamber with a base pressure of $\sim$4$\times$10$^{-10}$ Torr at room temperature. The ices are deposited onto a CsI window cooled to as low as $\sim$11 K using a closed-cycle Helium cryostat. These ices are grown using a 4.8 mm gas doser that is positioned close to the CsI subtrate at normal incidence. The temperature of the crystal is monitored and controlled using a LakeShore Model 335 controller with
two calibrated silicon diode sensors that have an estimated accuracy of 2 K and a relative uncertainty of 0.1 K. 
Transmission infrared spectra of the ices are obtained using a Fourier transform infrared (FTIR) spectrometer (Bruker Vertex 70v) with a resolution of 1 cm$^{-1}$ and with 60 scans taken per spectra.
Gas partial pressures were monitored during the diffusion experiments using a quadrupole mass spectrometer (Pfeiffer QMG 220M1). The desorbing molecules are monitored using a quadrupole mass spectrometer (Hiden IDP 300, model HAL 301 S/3) with a pinhole that is moved via a translational stage to $\sim$0.5 inches away from the ice. The experiments were performed using CO$_{2}$ gas (99.99 atom \% $^{12}$C, Sigma), $^{13}$CO$_{2}$ (99 atom \% $^{13}$C, $<$3 atom \% $^{18}$O, Sigma) and $^{13}$CO (99 atom \% $^{13}$C, $<$5 atom \% $^{18}$O, Sigma).

\subsection{Diffusion Experimental Procedures}\label{subsec:Diffproc}

\begin{figure*}[ht!]
	\plotone{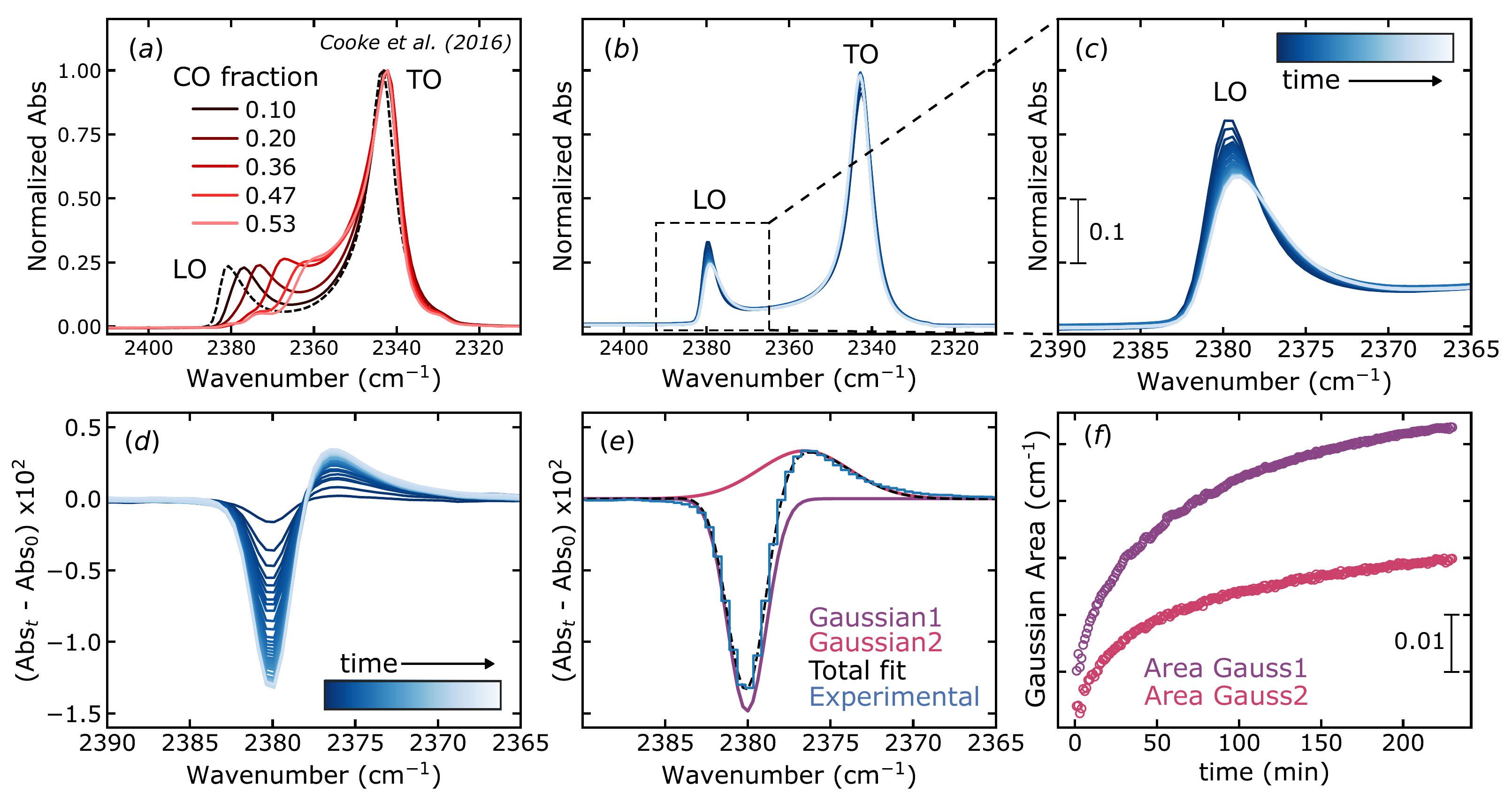}
	\caption{Strategy for fitting the changes in the CO$_2$ $\nu_3$ LO mode to determine the kinetics of CO diffusion into CO$_2$ ices $(a)$ Spectra of CO:CO$_2$ ice mixtures adapted from \citet{Cooke2016} showing the redshift in the CO$_2$ LO phonon mode with increasing CO ice fraction, the dashed line shows pure CO$_2$ ice for reference. $(b)$ Absorbance spectra of the CO$_2$ $\nu_3$ mode during CO diffusion into 30 ML of CO$_2$ at 20 K, $(c)$ shows a zoom in of the LO phonon mode for clarity. $(d)$ Subtraction spectra of the CO$_2$ LO phonon mode during CO diffusion at 20 K. $(e)$ An example fit to an experimental subtraction spectrum taken after CO has diffused into CO$_2$ for $\sim$200 minutes, the spectra are fit by optimizing the sum of the two gaussians. $(f)$ The resulting areas of the two gaussians plotted against the diffusion time.}
	\label{fig:examplefit}
\end{figure*}

The diffusion experiments consist of initially layered CO:CO$_2$ ices whose mixing is monitored using infrared spectroscopy. 
In each of these experiments $^{13}$CO and CO$_{2}$ were deposited sequentially at 11 K at a rate of $\sim$1 ML/minute to form the layered ice structures.  The deposited ice thicknesses were determined using IR absorption spectroscopy and Eq \ref{eq:abs}, which relates the column density to the ice absorbance:

\begin{equation} \label{eq:abs}
N_i(cm^{-2}) = \frac{\,cos\theta \int \tau_i(\nu) \, d\nu}{A_i} 
\end{equation}

where $N_i$ is the column density of the ice species $i$, $\theta$ is the angle of incidence between the IR field vector and the ice surface normal (here 45$^{\circ}$), $\int \tau_i(\nu) d\nu$ is the integrated area of the chosen IR band (in optical depth) and $A_i$ is the associated band strength adopted from \citet{Gerakines1995} and \citet{Bouilloud2015}. The column densities, $N_i$, were then converted to thicknesses in monolayers assuming 10$^{-15}$ molecules/ML, or to nanometers using the mass densities of CO and CO$_2$ ice from \citet{Satorre2008} and \citet{Roux1980}.
Following deposition, the layered ices were kept at 11 K for $\sim$10 minutes and were subsequently heated at 5 K minute$^{-1}$ to the desired temperature and maintained there for 2--4 hours. Time zero was taken when the isothermal temperature was reached. Infrared scans were taken every minute to monitor the ice composition.\\
\indent The different families of experiments are illustrated in Figure  \ref{fig:scheme}. The target and actual layer thicknesses, as well as the temperature at which mixing was monitored are listed in Table \ref{tab:spectro}. The fiducial experiment consisted of ices with target thicknesses of 30 ML $^{13}$CO followed by 30 ML of CO$_2$ and held at 20 K. We then carried out a series of experiments at different temperatures and with different ice thicknesses as well as experiments with isotopically labeled layers in order to extract the barrier for diffusion and elucidate the diffusion mechanism. We ran the temperature dependent experiments with the 30 ML:30 ML composition from T = 18--23 K. Above 23 K non-negligible CO desorption occurs and the diffusion rate is so rapid that the fits have large uncertainties. Below 18 K the diffusion rate is too slow to measure during our experimental timescale. In addition to the temperature dependent experiments, we also ran diffusion experiments for different CO:CO$_2$ thickness configurations at 20 K using the thicknesses shown in Figure \ref{fig:scheme}. \\

\begin{deluxetable*}{ccccccc}[ht!]
	\tablecaption{Initial ice thicknesses and diffusion temperatures used in the diffusion experiments and modeling. There are two experimental series aimed at elucidate temperature and ice thickness dependencies, as well as an isotopically labeled experimental series. \label{tab:spectro}}
	\tablecolumns{7}
	\tablenum{1}
	\tablewidth{0pt}
	\tablehead{
	    \colhead{Experiment} &
	    \colhead{Target Ice} &
		\colhead{T$_{diff}$} &
		\colhead{CO (1-0)  area} &
		\colhead{CO$_2$ $\nu_{3}$ area} &
		\colhead{CO thickness\tablenotemark{a}} &
		\colhead{CO$_2$ thickness\tablenotemark{a}} \\		
		\colhead{} & \colhead{(ML)} & \colhead{(K)} & \colhead{(cm$^{-1}$)} & \colhead{(cm$^{-1}$)} &  \colhead{$d$ (nm)} & \colhead{$h-d$ (nm)}
	}
	\startdata	
	\hline
	Temperatures & CO:CO$_2$ & & & & & \\
	\hline
	1 & 30:30 & 18 & 0.27 & 1.39 & 16 & 14\\
	2 & 30:30 & 19 & 0.27 & 1.60 & 16 & 16\\
	3 & 30:30 & 20 & 0.27 & 1.38 & 15 & 14\\
	4 & 30:30 & 21 & 0.26 & 1.45 & 15 & 14\\
	5 & 30:30 & 22 & 0.28 & 1.42 & 16 & 14 \\
 	6 & 30:30 & 23 & 0.27 & 1.47 & 15 & 15 \\
	\hline
	Ice thicknesses & CO:CO$_2$ & & & & &\\
	\hline
	3 & 30:30 & 20 & 0.27 & 1.38  & 15 & 14 \\
	7 & 10:50 & 20 & 0.09 & 2.30  & 5  & 23 \\
	8 & 50:10 & 20 & 0.45 & 0.50  & 26 & 5 \\
	9 & 20:20 & 20 & 0.13 & 0.78 & 8 & 8 \\
	\hline
	Isotope layers & CO:$^{i}$CO$_2$:$^{j}$CO$_2$\tablenotemark{b} & & & & & \\
	\hline
	10 & 30:15:15 & 20 & 0.28 & 0.77+0.70 & 16 & 8+7\\
	11 & 30:15:15 & 20 & 0.28 & 0.71+0.77 & 16 & 7+8 \\
	12 & 30:40:15 & 20 & 0.27 & 1.89+0.79 & 16 & 19+8 \\
	\hline
	\enddata
	\tablenotetext{a}{The uncertainty on the ice thickness in nanometers is estimated to be $\sim$15\%}
	\tablenotetext{b}{\textit{i} and \textit{j} refer to carbon mass 13-Carbon or 12-Carbon.}
	\tablecomments{We use the following band strengths and ice densities to calculate the CO and CO$_2$ thicknesses: A$_{^{12}CO_2}$($\nu_3$) = 1.1 $\times$10$^{-16}$, A$_{^{13}CO_2}$($\nu_3$) = 1.15 $\times$10$^{-16}$, A$_{^{13}CO}$(1-0)= 1.7 $\times$10$^{-17}$ cm molecule$^{-1}$ from \citet{Gerakines1995} and corrected for denisity in \citet{Bouilloud2015}, $\rho_{CO_2}$ = 1.1 g/cm$^3$ \citep{Satorre2008}, $\rho_{CO}$ = 0.8 g/cm$^3$ \citep{Roux1980}}
\end{deluxetable*}

\subsection{TPD Experimental Procedures}\label{subsec:TPDproc}

Temperature programmed desorption experiments are used to obtain the desorption energy of $^{13}$CO from CO$_{2}$ ice. Ices are grown using the same conditions described in section \ref{subsec:Diffproc}. In each experiment we first deposited $\sim$50 ML of CO$_2$ followed by $\leq$1 ML of CO. The CO$_{2}$ ice substrates were deposited at 11, 21, 23, 25, 40 and 50 K to obtain different CO$_2$ ice structures; the ice deposited at the lowest temperature is expected to be the most porous. Following CO$_2$ deposition the ice was cooled down to 11 K before depositing $^{13}$CO. The ices were heated at a constant rate of 1 K minute$^{-1}$. We subtract the mass background for $^{13}$CO and normalize the integrated QMS signal to the amount of CO deposited using the infrared spectra taken prior to heating.

\begin{deluxetable*}{ccCccccc}[b!]
	\tablecaption{CO:CO$_2$ diffusion experiments grouped by experiment type, together with the final fitted LO Gaussian areas, mixing rate coefficients and Fickian diffusion coefficients. \label{tab:resultsdiff}}
	\tablecolumns{8}
	\tablenum{2}
	\tablewidth{0pt}
	\tablehead{
		\colhead{Experiment} &
		\colhead{Target Ice} &
		\colhead{T$_{diff}$} & \colhead{LO$_{2381}$ area$_{f}$\tablenotemark{a}} & \colhead{LO$_{2375}$ area$_{f}$\tablenotemark{a}} & \colhead{k$_{mix}$\tablenotemark{b}} & \colhead{D$_{Fickian}$\tablenotemark{c}} \\
		\colhead{} & \colhead{(ML)} &
		\colhead{(K)} & \colhead{(cm$^{-1}$)} & \colhead{(cm$^{-1}$)}  &\colhead{(s$^{-1}$)} & \colhead{(cm$^2$ s$^{-1}$)} 
	}
	\startdata
	Temp dep & CO:CO$_2$ &  & & & &  \\
	\hline
	1 & 30:30 & 18 & 0.027 & 0.012  & 5.9 $\pm$ 2.0 $\times$10$^{-5}$ & 9.0 $\pm$ 0.6 $\times$10$^{-17}$\\
	2 & 30:30 & 19 & 0.043 & 0.026  & 2.0 $\pm$ 0.6 $\times$10$^{-4}$ & 2.3 $\pm$ 1.2 $\times$10$^{-16}$\\
	3 & 30:30 & 20 & 0.041 & 0.021  & 3.1 $\pm$ 1.0 $\times$10$^{-4}$& 2.5 $\pm$ 1.4  $\times$10$^{-16}$\\
	4 & 30:30 & 21 & 0.041 & 0.026  & 1.2  $\pm$ 0.4 $\times$10$^{-3}$ & 8.3 $\pm$ 0.5 $\times$10$^{-16}$\\
	5 & 30:30 & 22 & 0.051 & 0.028  & 3.2 $\pm$ 1.6 $\times$10$^{-3}$ & 2.4  $\pm$ 2.0 $\times$10$^{-15}$\\
	6 & 30:30 & 23 & 0.050 & 0.030  & 6.7 $\pm$ 3.4 $\times$10$^{-3}$  & 3.2 $\pm$ 2.8 $\times$10$^{-15}$\\
	\hline
	Ice thickness & CO:CO$_2$ & & & &  \\
	\hline
	3 & 30:30 & 20 & 0.041 & 0.021  & 3.1 $\pm$ 1.0 $\times$10$^{-4}$ & 2.5 $\pm$ 1.3 $\times$10$^{-16}$\\
	7 & 10:50 & 20 & 0.071 & 0.049  & 2.3 $\pm$ 1.4 $\times$10$^{-4}$ & 2.8 $\pm$ 5.8 $\times$10$^{-17}$\\
	8 & 50:10 & 20 & 0.010 &  **  & 3.7 $\pm$ 1.1 $\times$10$^{-3}$ & 1.3 $\pm$ 6.8 $\times$10$^{-15}$\\
	9 & 20:20 & 20 & 0.016 & 0.004 & 3.8 $\pm$ 1.2 $\times$10$^{-4}$ & 1.1 $\pm$ 2.3 $\times$10$^{-16}$ \\
	\hline
	Isotopic layers & CO:$^{i}$CO$_2$:$^{j}$CO$_2$\tablenotemark{d} & & \textit{i:j} & \textit{i:j}  &  \\
	\hline
	10 & 30:15:15  & 20 & 0.012:0.027 & 0.009:0.011 & 5.7 $\pm$ 1.8 $\times$10$^{-4}$\tablenotemark{*} & 5.8 $\pm$ 0.4$\times$ 10$^{-16}$\tablenotemark{*}\\
	11 & 30:15:15  & 20 & 0.012:0.028 &0.007:0.016  & 4.8 $\pm$ 1.5 $\times$10$^{-4}$\tablenotemark{*} & 4.7 $\pm$ 2.9 $\times$10$^{-16}$\tablenotemark{*}\\
	12 & 30:40:15  & 20 & 0.058:0.025 & 0.040:0.008 & 3.0 $\pm$ 0.9 $\times$10$^{-4}$\tablenotemark{*} & 5.1 $\pm$ 3.0 $\times$10$^{-16}$\tablenotemark{*} \\
	\hline
	\hline
	\enddata
	\tablenotetext{a}{Area of Gaussian fit at the end of the diffusion experiment period}
	\tablenotetext{b}{Mixing rate calculated by fitting equation \ref{eq:5} to the experimental data.}
	\tablenotetext{c}{Fickian diffusion coefficient found by fitting \ref{eq:4} to the experimental data}
	\tablenotetext{d}{\textit{i} and \textit{j} refer to carbon mass 13-Carbon or 12-Carbon.}
	\tablenotetext{*}{Calculated for the total CO$_2$ ice thickness by summing together the two layers.}
	\tablecomments{** The integrated Gaussian area was too low to obtain a good fit.}
\end{deluxetable*}

\section{Diffusion Analysis and Results}\label{sec:Diffresults}

\begin{figure*}[t!]
	\plotone{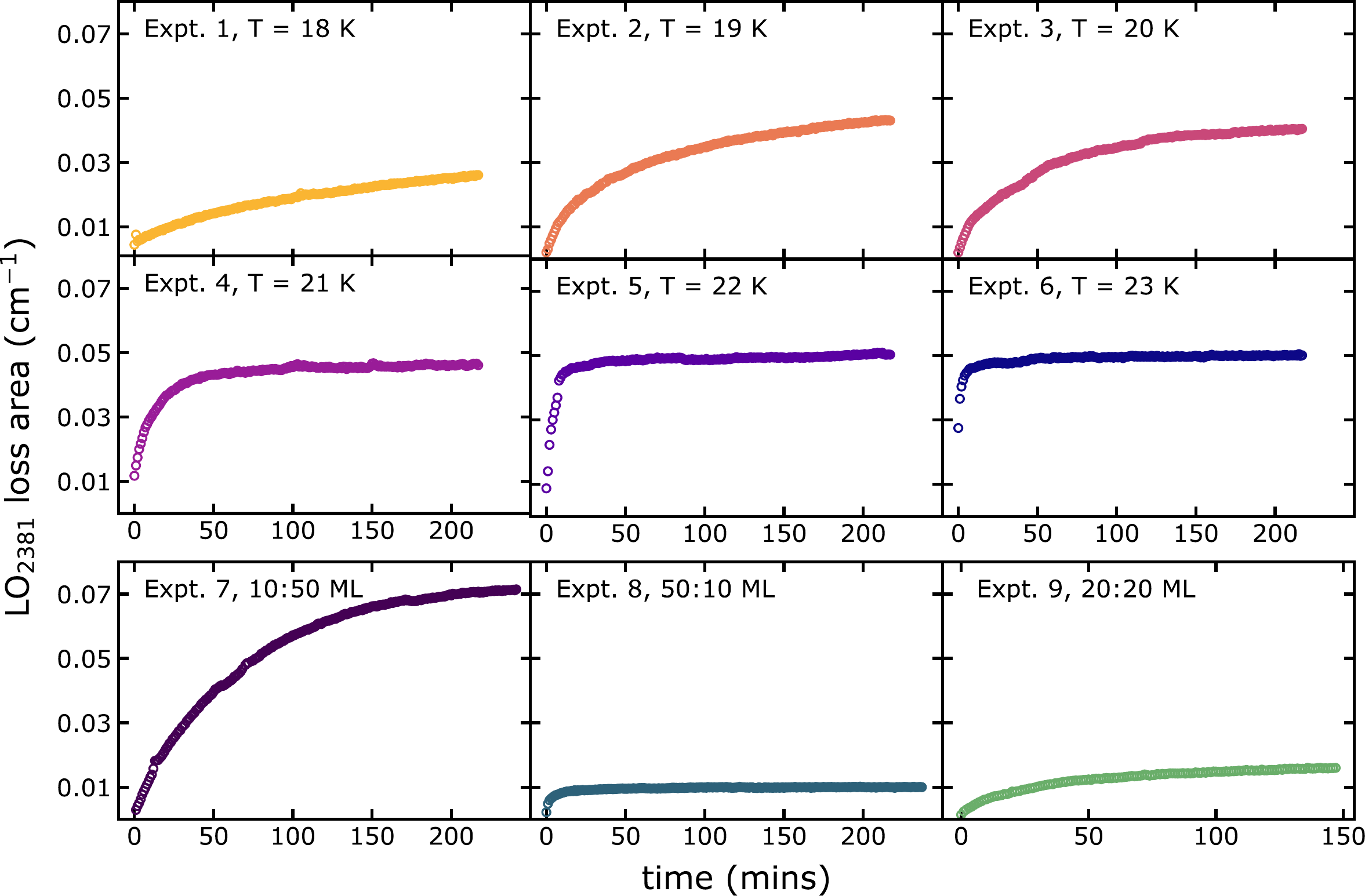}
	\caption{Experimental kinetic curves for the diffusion of CO into CO$_2$ as traced by the CO$_2$ $\nu_3$ LO mode. Here, the area of the Gaussian corresponding to the loss of the pure CO$_2$ environment upon mixing with CO is plotted against time. The top two rows show the results of experiments conducted for 30 ML CO: 30 ML CO$_2$ ices at six temperatures between 18--23 K. The bottom row shows the results for diffusion experiments where the CO:CO$_2$ thickness ratio is 1:5, 5:1 and 1:1.}
	\label{fig:resultsdiff}
\end{figure*}

In this section, we describe the results and analysis of the CO:CO$_2$ diffusion experiments. In section \ref{subsec:spectral}  we present the spectral analysis used to follow CO diffusion into the CO$_2$ based on changes in the $\nu_3$ LO phonon mode. In section \ref{subsec:diffresults} we describe the outcome of all bi-layered diffusion experiments using the spectral analysis from section \ref{subsec:spectral}. In section \ref{subsec:model} we describe the model framework used to quantify the diffusion rate in each experiment. In section \ref{subsec:barrier} we apply the models to the experimental data and extract the CO-CO$_2$ diffusion barrier. Finally, in section \ref{subsec:isotope} we present the results of experiments in which isotopically labelled CO$_2$ layers are employed to further constrain the diffusion mechanism. 

\subsection{Spectral Analysis}\label{subsec:spectral}

During the isothermal diffusion experiments we monitor changes in the CO$_2$ LO phonon mode. Figure \ref{fig:examplefit}(a) shows spectra of CO:CO$_2$ ice mixtures of various CO concentration, reproduced from \citet{Cooke2016}; the CO$_2$ LO phonon mode is perturbed when CO$_2$ is mixed with CO and thus can be used as a tracer of CO diffusion in CO$_2$ ices. \\
\indent Figure \ref{fig:examplefit}(b)-(c) shows an example of the CO$_2$ LO phonon mode during the diffusion experiment at 20 K. The LO phonon frequency at $t=0$ is taken as reference for the pure CO$_2$ ice and is $\sim$2381 cm$^{-1}$ (4.2 $\mu$m). With time, we observe a decrease in the LO phonon intensity at 2381 cm$^{-1}$ and an apparent broadening of the feature towards lower frequencies. The subtraction spectra (Figure \ref{fig:examplefit}(d)) reveal two distinct features: a loss centered at $\sim$2381 cm$^{-1}$ and a growth centered around 2375 cm$^{-1}$. Two Gaussians are fit to the subtraction spectra (Figure \ref{fig:examplefit}(e)) and their sum is optimized in Python using the \textit{scipy.optimize.nnls} optimization package. The resulting negative Gaussian (Gaussian 1) is considered as a loss of the original pure CO$_2$ ice environment, while the positive Gaussian (Gaussian 2) arises from the new CO-CO$_2$ mixed environment. The new redshifted LO feature, while growing in intensity during diffusion experiment, does not shift in frequency, which can be contrasted to CO:CO$_2$ mixtures we deposited from gases in \citet{Cooke2016}, where the frequency changed with mixture concentration.\\
\indent While both Gaussians can be used to model mixing of CO into the CO$_2$ layer, we use the negative Gaussian 1 to extract the diffusion coefficients; the integrated area of Gaussian 1 is larger than that of Gaussian 2, allowing us to better fit the fast mixing kinetics within the first 10 minutes of the diffusion experiments. 

\subsection{CO diffusion experimental results}\label{subsec:diffresults}

Figure \ref{fig:resultsdiff} and Table \ref{tab:resultsdiff} show the results of the diffusion experiments. The top two rows of Figure \ref{fig:resultsdiff} show the outcome of diffusion experiments with close to identical ice thicknesses but run at six different temperatures between 18 and 23 K. At 18 K mixing is not complete after 250 min, while at 23 K it is complete within the first $\sim$10 min. The final mixing fraction, as traced by the loss of the LO mode, is almost constant above 18 K i.e. the ice morphology is almost independent of ice temperature within the explored range; by contrast the loss rate depends strongly on temperature. This is the expected behavior for a system in which diffusion is driven by the random movement of the more volatile species, motivating our model choices below. \\ 
\indent The bottom row of Figure \ref{fig:resultsdiff} shows the experiments for different CO:CO$_2$ thicknesses; Experiment 7: 10 ML CO and 50 ML CO$_2$, Expt 8: 50 ML CO and 10 ML CO$_2$ and Expt 9: 20 ML CO and 20 ML CO$_2$. The final mixed fraction depends on the CO$_2$ ice thickness as expected, i.e. thicker CO$_2$ ices can host more CO molecules. We explore the dependence of the diffusion rate on ice thickness and CO:CO$_2$ ratio quantitatively in section \ref{subsec:barrier}.

\subsection{Fickian diffusion modeling}\label{subsec:model}

We use a Fick's second law model to extract the diffusion coefficients and barrier for CO diffusion into CO$_2$ ice. Fick's law has been applied by \citet{Karssemeijer2014}, \citet{Mispelaer2013} and \citet{Lauck2015} to model CO diffusion in amorphous solid water (ASW) ices.This law should apply if the ice mixing is dominated by random walk diffusion of the more volatile CO into the CO$_2$ matrix resulting in a concentration gradient across the ice depth. We also fit the kinetic data with exponentials to give a rate coefficient and time associated with CO mixing into the CO$_2$ layer.

\begin{figure*}[ht!]
	\plotone{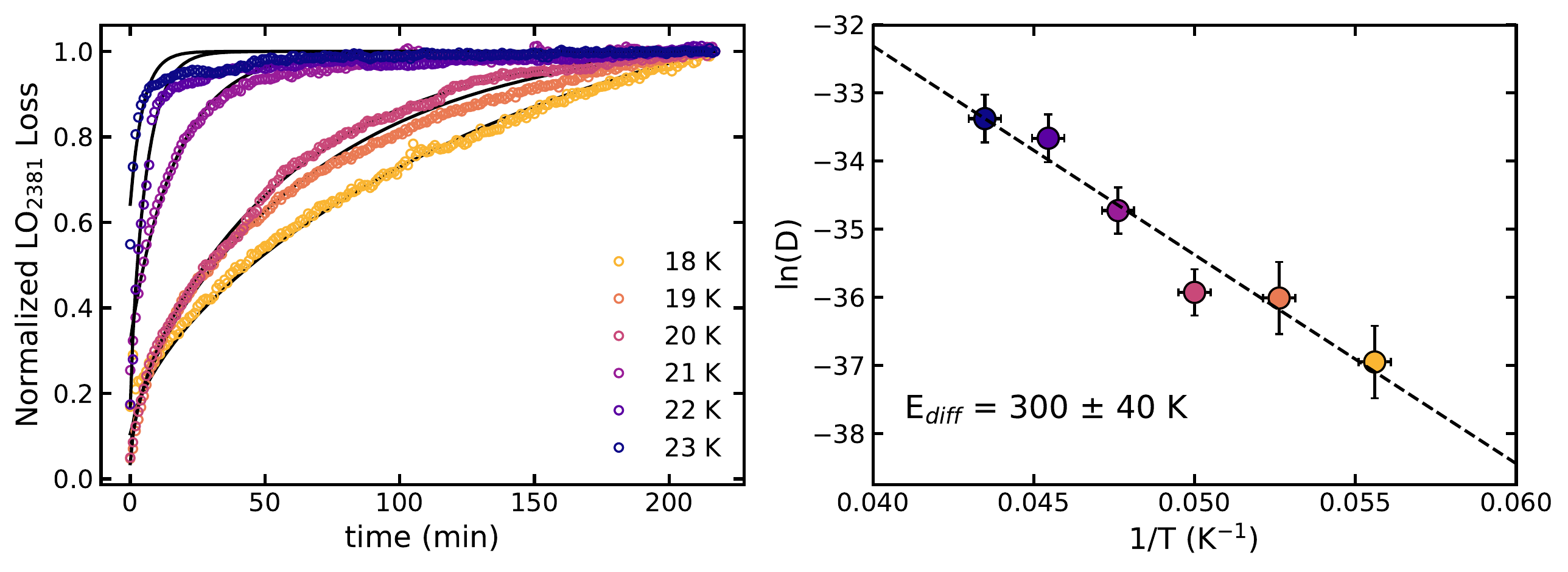}
	\caption{Temperature dependent kinetics of the CO diffusion into CO$_2$ ice fit using Fick's second law (left) along with the Arrhenius Law plot (right) for the temperature dependent diffusion coefficients}
	\label{fig:arrhenius}
\end{figure*}

We adopt a Fickian diffusion model modified from \citet{Lauck2015} and \cite{Bergner2016}.  
The general form of Fick's second law for a 1-D system is:

\begin{equation}
\frac{\partial c(z,t)}{\partial t} =  D(T) \frac{\partial^2 c(z,t)}{\partial z^2}
\end{equation}

\indent where $c(z,t)$ is the concentration of the diffusant CO as a function of time, $t$, and position, $z$, and $D(T)$ is the temperature dependent diffusion coefficient. In the layered CO:CO$_2$ system we define $z=0$ as the substrate height, $z=d$ as the interface height between CO and CO$_2$ layers, and $z=h$ as the vacuum interface. To calculate the height of the CO/CO$_2$ and vacuum interfaces, we use densities of 1.1 g/cm$^3$ \citep{Satorre2008} and 0.8 g/cm$^3$ \citep{Roux1980} for CO$_2$ and CO, respectively.  For a system where CO desorption is negligible we impose boundary conditions such that the flux of CO at the CsI subtrate and at the vacuum interface is zero, i.e. $\frac{\partial c(z,t)}{\partial t} = 0 $ at $z=0$ and $z=h$. At $t=0$, we assume the concentration of CO is $c_0$ in the CO layer and zero everywhere else. Applying these boundary conditions gives a general solution that may be integrated to find the amount of CO in the CO$_2$ layer. Dividing this through by the total amount of CO gives a mixed fraction, $N_{mix}$:

\begin{eqnarray}
N_{mix}(t) = \frac{1}{dc_0} \int_{d}^{h} c(z,t) dz = \frac{h-d}{h} - \\ \nonumber \sum_{n=1}^{\infty} \frac{2h}{n^2\pi^2d} sin^2\left(\frac{n\pi d}{h}\right) \, exp\left(-\frac{n^2\pi^2}{h^2}Dt\right)
\end{eqnarray}

This is adjusted to account for mixing during the fast temperature ramp by using a time offset, $t_0$, and for uncertainties in the measured ice thickness using a nuisance parameter, $N_0$, yielding  

\begin{eqnarray} \label{eq:4}
N_{mix}(t) = N_0 \frac{h-d}{h} - \sum_{n=1}^{\infty} \frac{2N_0h}{n^2\pi^2d} sin^2\left(\frac{n\pi d}{h}\right) \\ \nonumber 
\times \, exp\left(-\frac{n^2\pi^2}{h^2}D(t+t_0)\right)
\end{eqnarray}

Here $D$, $N_0$ and $t_0$ are free parameters that are fit to the experimental mixing fraction of CO over time. We use the python non-linear least squares routine \textit{scipy.optimize.curve\_fit} to fit equation \ref{eq:4}.  \\
We also fit the mixing of CO into CO$_2$ using exponentials. Fitting exponentials to the data allows us to directly extract a time constant associated with the diffusion process. A similar method has been used to fit the kinetics of molecules diffusing into ASW ice \citep{Mispelaer2013}. The exponential equation describing the time dependent mixed fraction, $N_{mix}$, is:

\begin{equation} \label{eq:5}
N_{mix}(t) = N_0 \, e^{-\left(k_{mix}t\right)^n}
\end{equation}

where $k_{mix}$ is the mixing rate coefficient in $s^{-1}$ and $n$ is the kinetic order. 
The diffusion coefficient, D, can be roughly approximated from the mixing rate coefficient using Einstein's relationship:

\begin{equation}\label{eq:6}
D \simeq \frac{k_{Av} \, (h-d)^2}{2}
\end{equation}

\indent Where $h-d$ is the thickness of the CO$_2$ ice in which the CO diffuses. This equation generally gives us the same order of magnitude diffusion coefficients as obtained using the Fickian model.

\subsection{Diffusion Kinetics and Barriers}\label{subsec:barrier}

Consistent with the qualitative analysis above the Fickian diffusion coefficients increase with temperature from $\sim$1x10$^{-16}$ cm$^{2}$ s$^{-1}$ at 18 K to $\sim$3x10$^{-15}$ cm$^{2}$ s$^{-1}$ at 23 K. The mixing rates obtained from exponentials fits to the kinetic data likewise increase with temperature and are shown in Table \ref{tab:resultsdiff}. There are two major sources of uncertainty on the diffusion coefficients that are propagated into the uncertainty on the diffusion barrier. At the higher temperatures (T$>$20 K), the largest source of uncertainty is the choice of the $t=0$ point, which can change the diffusion coefficient by up to 50\%. The largest source of uncertainty for the experiments where T$\leq$20 K arises from the thickness determination and is a combination of uncertainties in the CO and CO$_2$ band strengths and their densities. \\
\indent A weighted linear regression to the Fick's Law Arrhenius plot (Figure \ref{fig:arrhenius}) yields a diffusion energy barrier of 300 $\pm$ 40 K. We also fit an Arrhenius Law to the mixing rate coefficients (not shown here) and find a barrier of 380 $\pm$ 30 K, indicating that the derived barrier is robust to the choice of model.\\
\indent Comparing the ices with different thickness configurations we find that the mixing timescale decreases with decreasing CO$_2$ ice thickness. Based on the exponential fit, the characteristic mixing time constant is 4.5 minutes for the thin 10 ML CO$_2$ ice and 72 minutes for mixing into the thick 50 ML CO$_2$ ice. 
Likewise, the Fickian diffusion coefficients increase from 3 $\times$10$^{-17}$ cm$^{2}$ s$^{-1}$ in the 10 ML CO: 50 ML CO$_2$ ice to 1 $\times$10$^{-15}$ cm$^{2}$ s$^{-1}$ in the 50 ML CO: 10 ML CO$_2$ ice, i.e. it is larger for larger CO:CO$_2$ ratios. 
The diffusion rates extracted for experiments with the same CO:CO$_2$ thickness ratio but different total ice thicknesses (30 ML:30 ML and 20 ML:20 ML) are the same within experimental error, indicating that the CO:CO$_2$ thickness ratio together with temperature control the diffusion rate, and that total ice thickness is not an important factor. \\

\subsection{Isotopic Studies}\label{subsec:isotope}

\begin{figure*}[ht!]
	\includegraphics[height=5 cm]{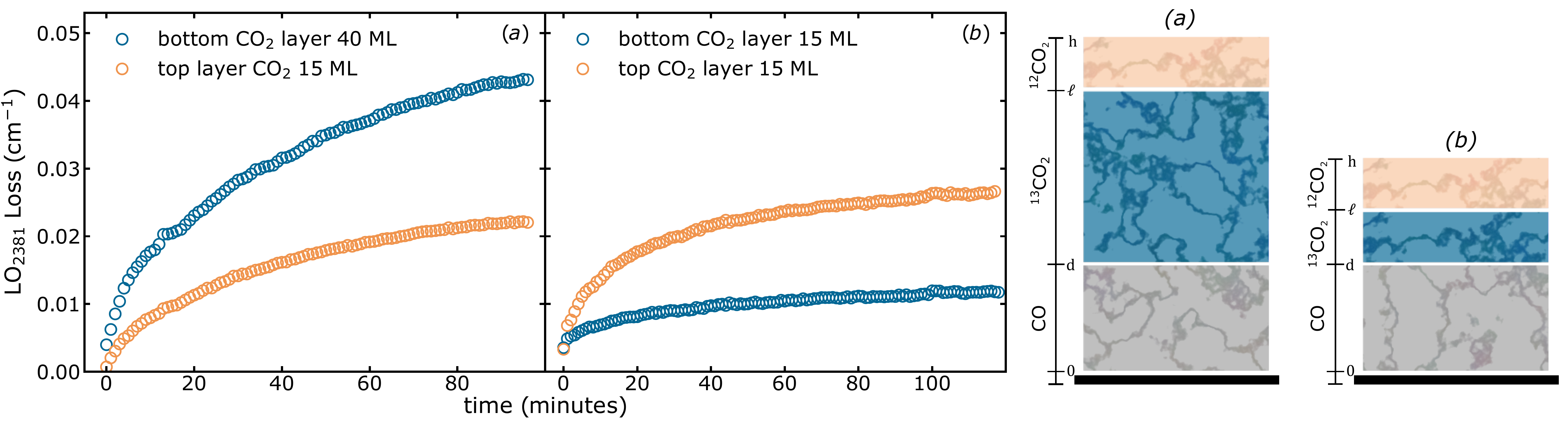}
	\caption{Experimental CO diffusion into isotopic layers of CO$_2$. Panel \textit{(a)} shows diffusion of CO into a 40 ML $^{13}$CO$_2$: 15 ML $^{12}$CO$_2$ ice. Panel \textit{(b)} shows CO diffusion into two 15 ML thick $^{13}$CO$_2$: $^{12}$CO$_2$ isotopic layers.}
	\label{fig:isotopes}
\end{figure*}

The isotopically labeled layered ice experiments provide further insight into the mechanism of CO:CO$_2$ diffusion. Experiments were conducted with layered isotopic CO$_2$ ices as shown in Figure \ref{fig:scheme} and Table \ref{tab:spectro}. We test two different isotopic thickness configurations in which we layered 30 ML of CO with (1) 40 ML $^{13}$CO$_2$ then 15 ML $^{12}$CO$_2$ (Fig \ref{fig:isotopes}(a)) and (2) 15 ML $^{13}$CO$_2$ then 15 ML $^{12}$CO$_2$ (Fig \ref{fig:isotopes}(b)).\\
\indent Figure \ref{fig:isotopes} shows the kinetic curves for CO mixing into the two CO$_2$ layers. In Fig \ref{fig:isotopes}(a) we see that the middle 40 ML layer has a larger final LO Gaussian loss area, corresponding to a larger number of mixed CO molecules, than the top 15 ML layer. By contrast, in Fig \ref{fig:isotopes}(b) the top 15 ML is able to host more CO than the bottom 15 ML despite their equal thicknesses. We also switched the isotopic order and layered 15 ML $^{12}$CO$_2$ then $^{13}$CO$_2$ and found that the top 15ML layer always hosts more CO regardless of the order of the two isotopologues. In each of these three experiments, we found that the final mixed fraction in the top 15 ML layer was the same. We discuss the physical interpretation  of these results further in section \ref{sec:Discuss}. \\
\indent We calculate the diffusion coefficients for CO through the total ice thickness by summing the LO loss feature in both layers, reported in Table \ref{tab:resultsdiff}. We do this purely to check whether the diffusion coefficients calculated from the summing the two 15 ML layers are the same as the 30 ML experiment at 20 K and we find that two are indeed the same within experimental uncertainties.

\section{TPD Analysis and Results}\label{sec:TPDresults}

\subsection{TPD Analysis}

\begin{figure*}[ht!]
	\plotone{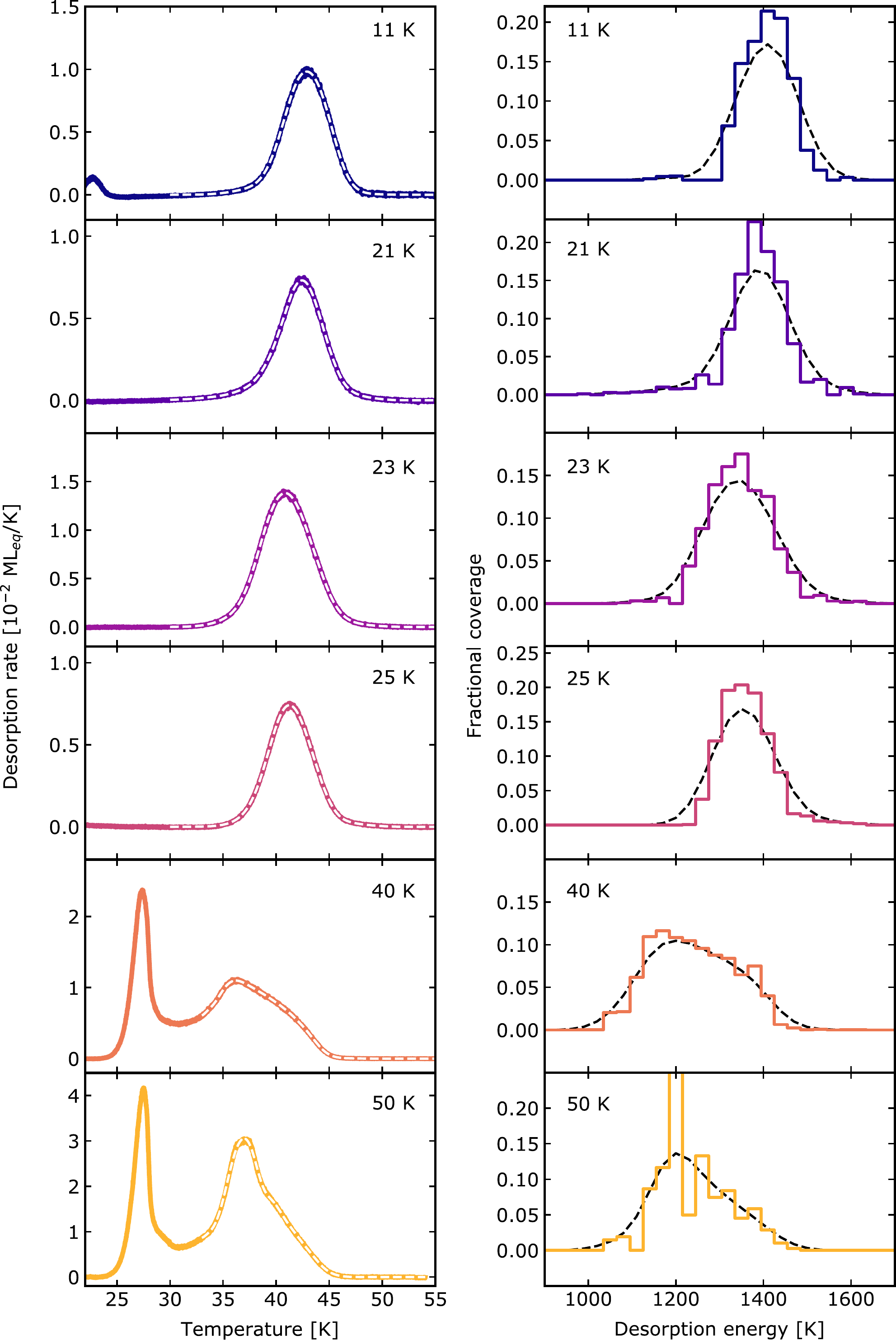}
	\caption{$^{13}$CO temperature programmed desorption curves (\textit{left}) and related desorption energy distributions (\textit{right}) from CO$_2$ ices at various deposition temperatures listed in table \ref{tab:Edes}. The TPD spectrum of CO desorption from CO$_2$ ice grown at 11 K also displays a small peak below 25 K that we attribute to CO co-desorption with hydrogen that is deposited from the chamber background.}
	\label{fig:TPD}
\end{figure*}

The TPD traces for CO desorbing from CO$_2$ ices are shown in the left panel of Figure \ref{fig:TPD}. The spectra display one or two peaks between 25--50 K, depending on the ice surface area and corresponding CO coverage. CO desorption from CO$_2$ ices deposited at 40 and 50 K display two TPD peaks in this temperature regime. The first peak corresponds to multilayer CO desorption and occurs around 28 K, consistent with previous measurements in the literature \citep{Oberg2005,Noble2012,Collings2015,Fayolle2016}. The higher temperature peak is associated with submonolayer desorption of CO from the CO$_2$ ice surface. This peak is broader than in the multilayer regime, indicating a larger range of binding sites and associated energies, even when the CO$_2$ ice is quite compact. \\
\indent CO$_2$ ices deposited at temperatures $\leq$25 K have a single desorption peak associated with sub-monolyer CO desorption from the surface of the CO$_2$ ices. An additional desorption peak is seen near the CO$_2$ desorption temperature (not shown here), probably due to CO diffusion into the CO$_2$ pores and subsequent entrapment due to pore collapse. Similar entrapment has been seen for other volatile species within porous ASW ices \citep{Collings2003,Fayolle2011,MartinDomenech2014}. No multilayer peak is observed for these ices, which is consistent with the expectation that ices deposited at lower temperatures are more porous and therefore present a larger surface for adsorbing molecules.

\begin{deluxetable}{cccc}[b!]
	\tablecaption{CO column densities, mean desorption energies and full-width half maxima for submonolayer CO desorption from CO$_2$ ices at different deposition temperatures.\label{tab:Edes}}
	\tablecolumns{4}
	\tablenum{3}
	\tablewidth{0pt}
	\tablehead{
		\colhead{CO$_2$ Temp} &
		\colhead{CO column density} &
		\colhead{E$_{des}$} &
		\colhead{FWHM} \\
		\colhead{(K)} &
		\colhead{(10$^{15}$ molecules/cm$^2$)} &
		\colhead{(K)} &
		\colhead{(K)}
			}
	\startdata
	11 & 1.0 & 1407 & 71 \\
	21 & 0.7 & 1385 & 86 \\
	23 & 0.8 & 1347 & 84 \\
	25 & 0.5 & 1361 & 73\\
	40 & 0.8 & 1240 & 105 \\
	50 & 0.8 & 1239 & 94\\
	\enddata
\end{deluxetable}

\subsection{Desorption Barriers} \label{subsec:desbarrier}

\begin{figure}[t!] 
	\includegraphics[width = 8.5 cm]{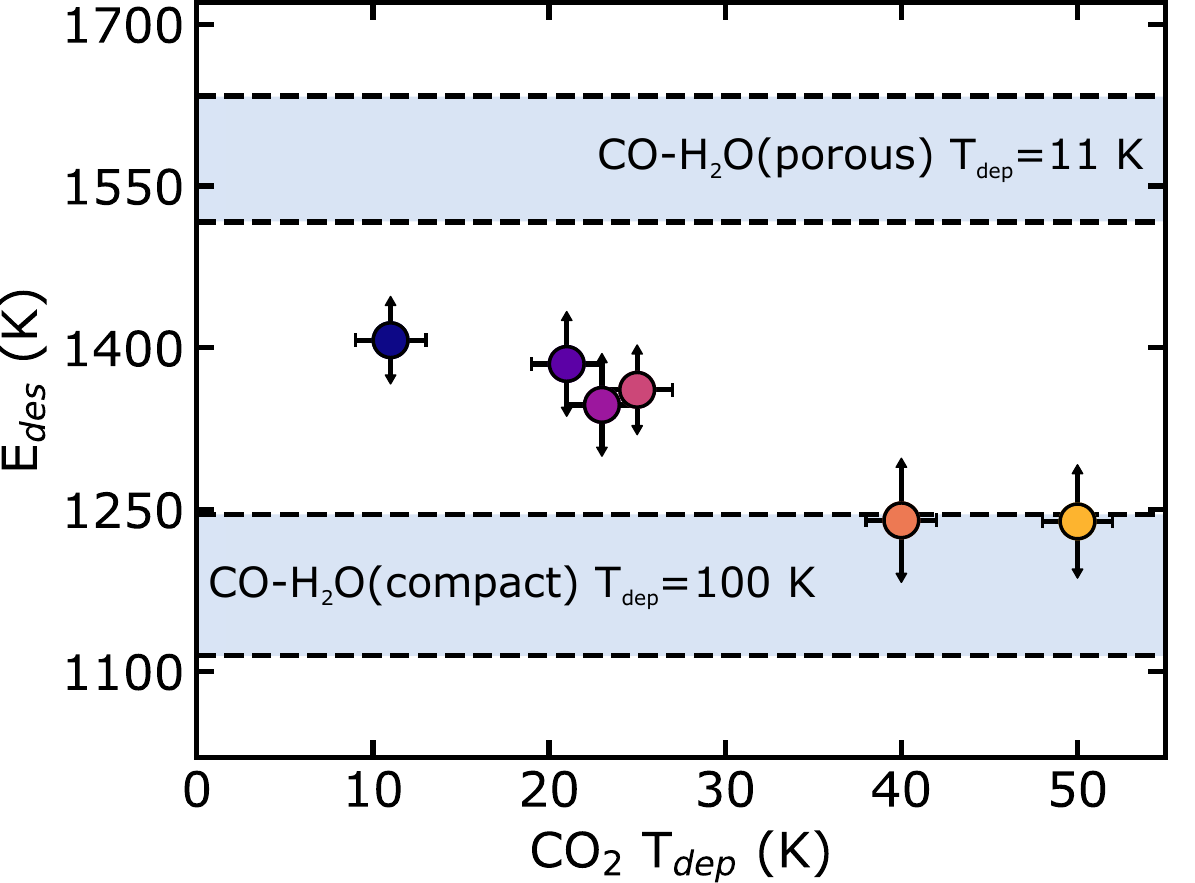}
	\caption{Desorption energies of $^{13}$CO from CO$_2$ ice for various CO$_2$ deposition temperatures. The blue panels show the average desorption energy with FWHM distributions for porous H$_2$O ice deposited at 11 K and for compact H$_2$O ice deposited at 100 K reported in \citet{Fayolle2016}}
	\label{fig:edes}
\end{figure}

Figure \ref{fig:TPD} shows the TPD curves for CO desorption from CO$_2$ ices deposited at 11, 21, 23, 25, 40 and 50 K. The TPD curves are fit using the Polanyi-Wigner equation:

\begin{equation}
- \frac{d\theta}{dT} = \frac{\nu}{\beta} \,\theta^n \, e^{-E_{des}/T}
\end{equation}

where $\theta$ is the CO ice coverage, $T$ is the temperature in K, $\nu$ is a pre-exponential frequency factor in s$^{-1}$, $\beta$ is the heating ramp rate in K s$^{-1}$, $n$ is the desorption order and $E_{des}$ is the desorption energy in K. 
We see that the peak desorption temperature increases as the CO$_2$ deposition temperature is decreased. The trailing edges of the sub-monolayer CO peaks also extends to higher temperature for more porous CO$_2$ ices deposited at lower temperatures. \\
\indent To derive the desorption energy distribution for CO on CO$_2$ we fit the TPD traces with a linear combination of first order kinetics using the methods of \citet{Doronin2015} and described in detail in \citet{Fayolle2016}. We use an energy step interval of 30 K to fit desorption kinetics between 900 and 1800 K. The resulting desorption energy distributions are shown in the right panel of Figure \ref{fig:TPD}.\\ \indent The mean desorption energies and desorption energy distributions, defined by the peak FWHM, for the various CO$_2$ deposition temperatures are shown in Figure \ref{fig:edes} and in Table \ref{tab:Edes}. The desorption energies for CO from CO$_2$ ices increase with decreasing deposition temperature and range from 1239 K to 1407 K for CO$_2$ ices deposited at 50 K and 11 K respectively. We show the desorption energies of CO from H$_2$O ice deposited at 11 K (porous) and at 100 K (compact) from \citet{Fayolle2016} for comparison. The increase in CO-CO$_2$  desorption energy with decreasing CO$_2$ deposition temperature is likely due to an increase in porosity and therefore number of strongly bound sites. In the submonolayer regime, mobile molecules tend to fill the deeper adsorption sites, resulting in a shift in the mean E$_{des}$ to higher energies (e.g. \citet{Fillion2009}).\\
\indent In the experiments where CO$_2$ is deposited at temperatures between 11--25 K, CO is slowly out-gassing between the CO and CO$_2$ desorption peaks, probably due to a combination of slow CO diffusion and CO$_2$ ice rearrangement during the TPD warm-up. CO out-gassing is the largest for the experiment where CO$_2$ ice was deposited at 11 K, consistent with expectations that this ice has the highest CO$_2$ porosity and therefore highest CO trapping efficiency. To avoid including the slow out-gassing effect into our calculation of the CO-CO$_2$ surface desorption energy, we fit a baseline to the TPD spectra before and after CO surface desorption, where the latter includes the spectral region where CO is slowly out-gassing. If the contribution from CO outgassing is instead included in the fit, the average desorption energies are systematically higher, but this increase is only significant for the 11 K CO$_2$ ice, where ignoring the baseline correction results in a $\sim$100 K increase in the desorption energy estimate compared to our reported value.

\section{Discussion}\label{sec:Discuss}

\subsection{Diffusion Mechanisms}

\begin{figure*}[ht!]
	\plotone{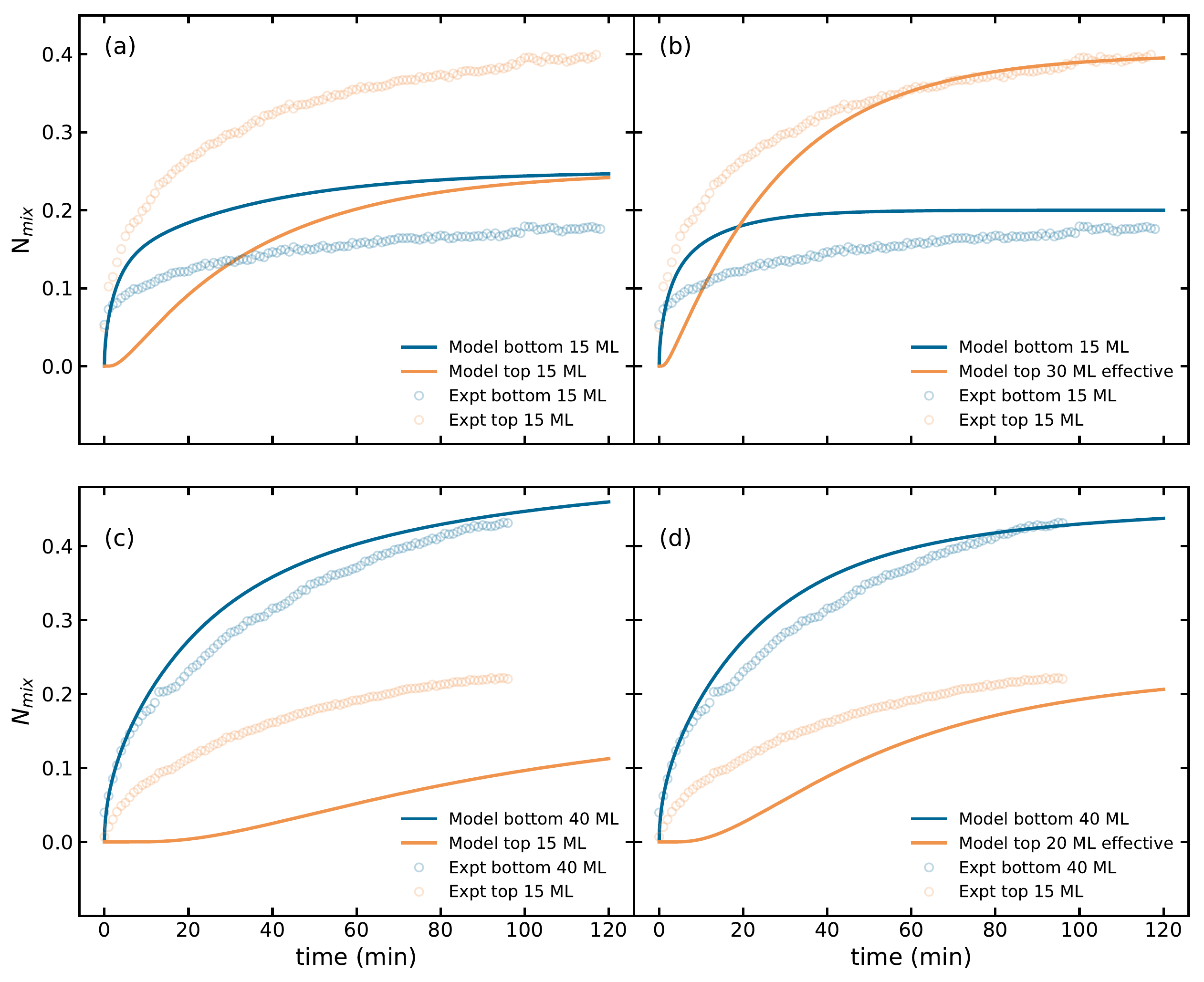}
	\caption{Fick's Law model for CO diffusion into isotopic layers of CO$_2$. Panels a) and c) model are modeled using the CO$_2$ thicknesses used in the experiment and a relevant diffusion coefficient from the measurements at 20 K. Panels b) and d) show adjustments to the Fickian model by scaling the heights to an ``effective'' thickness by assuming that the top layer has an additional surface area that is not present in the lower layer. The diffusion coefficient is increased to assume that once the CO reaches a pores in the lower CO$_2$ it can move more rapidly along the pore surface to the top layer. The faint open circles display the experimental kinetic traces for comparison.}
	\label{fig:modeliso}
\end{figure*}

\indent There are three main diffusion mechanisms proposed in the literature that are relevant for low temperature interstellar ices and their laboratory analogs: swapping of lattice molecules \citep{Oberg2009,Fuchs2009,Garrod2013}, movement into empty vacancy sites in the lattice (e.g. \citet{Lamberts2013,Lamberts2014,Chang2014} and surface hopping along adsorption sites in pores \citep{Garrod2013b}. The former two are \textit{bulk} diffusion processes, which are expected to have large barriers compared to pore surface hopping.\\
\indent Our experiments provide three different lines of evidence that, in the case of CO diffusion into CO$_2$, the main diffusion mechanism is that of surface diffusion in pores and on the ice surface: the magnitude of the extracted diffusion barrier, the evolution of the LO mode during diffusion, and the observed diffusion pattern through isotopically labeled ice layers.\\
\indent First, the low diffusion barrier and the low diffusion-desorption energy barrier ratio of 0.21-0.24 is consistent with surface diffusion, but not with models of bulk diffusion in which ratios ranging from 0.5 for diffusion by swapping \citep{Garrod2013} up to 1 for movement into interstitial sites \citep{Chang2014} have been employed. Similar low diffusion barriers have been measured for the diffusion of volatile species into porous water ices \citep{Mispelaer2013,Lauck2015}. \\
\indent Secondly, pore diffusion is the best explanation for how the CO$_2$ LO phonon mode evolves during the isothermal diffusion experiments. Our previous experiments reported in \citet{Cooke2016} have shown that bulk mixing of CO into CO$_2$ ices redshifts the CO$_2$ LO mode linearly with the concentration of CO. If diffusion occurred through swapping there should be a smooth change in the CO$_2$ ice lattice with time which should increasingly redshift the LO mode as more CO diffuses into the CO$_2$. This is not observed; rather, we observe the growth of a new feature at a single redshift. The diffusion mechanism then, does not change the bulk lattice structure during CO diffusion, and this is only consistent with either surface hopping or movement into interstitial sites.\\
\indent The third line of evidence comes from the behavior of the isotopically labeled CO$_2$ layered experiments in section \ref{subsec:isotope}. If diffusion occurs by the random walk into the CO$_2$ layer with a homogeneous distribution of binding sites, we would expect diffusion into the top layer should be delayed with respect to the bottom layer, and the final mixing fractions should be the same for two layers of the same thickness. In Figure \ref{fig:modeliso} we show a toy model to demonstrate this point. On the left-hand-side of Figure \ref{fig:modeliso} (panels (a) and (c)), we input our experimentally measured ice thicknesses into Fick's law to model the mixing of CO into the CO$_2$ isotopic layers and compare with the actual experiments. Contrary to model predictions, in our experiment the top isotopic layer hosts more CO molecules per ML of CO$_2$ compared the to the bottom CO$_2$ layer; the Fick's law model (solid lines) expects equal final mixing of CO into CO$_2$ layers of the same thickness. The second discrepancy is the predicted time delay for CO mixing into the top layer, which is not seen in our experimental data.\\
\indent One possible explanation for the discrepancy in the final mixing fractions is that the top layer has a larger number of surface sites, i.e. the binding sites for CO are not distributed homogeneously across the CO$_2$ ice height.
We incorporate this into the toy model by assigning the top layer an ``effective thickness" to best match the experimental mixing fractions by eye.
In the case of 15 ML:15 ML ice, we need to roughly double the thickness of the top layer to reproduce the relative mixing fractions seen in the experimental curves. In the 40 ML: 15 ML case we increase the top layer by around 33\%. In the pore-diffusion scenario, this effect can be explained physically by a larger number of surface sites per CO$_2$ ML in the top isotopic layer due to additional surface binding sites at the vacuum interface. Changing the effective thicknesses of the two layers does not resolve the above noted mismatch between predicted and experimental mixing delay times. The immediate appearance of mixing in the top ice layer (within the measurement time scale) is best explained by a rapid pore diffusion, which is faster than expected using our solution to Fick's law that. One possible explanation for the rapid pore-diffusion is that initially absorbed CO molecules at the CO-CO$_2$ boundary could facilitate faster diffusion of subsequent CO molecules via decreased van der Waals interactions. Because the CO-CO adsorption energy is lower than that of CO-CO$_2$ the diffusion kinetics would reflect the CO-CO self-diffusion barrier. This possibility has been suggested previously by \citet{Lauck2015} for the case of CO diffusing through porous water ices. \\
\indent In summary, diffusion of CO through CO$_2$ ice most likely occurs through internal pores; this theory is supported by the low diffusion-desorption energy barrier ratio, the evolution of the CO$_2$ $\nu_3$ LO mode during CO diffusion, and the surface accumulation of CO in the isotopically labeled ice experiments.\\

\subsection{CO$_2$ LO phonons for tracing diffusion}
We have presented a new method for studying diffusion processes in CO$_2$ bearing ices based on the sensitivity of the CO$_2$ LO phonon mode to the ice environment. Diffusion kinetics in ices have most commonly been measured via decreases in IR absorption of the diffusing molecule after desorption. Typically, a layered or mixed ice is heated temperatures above the desorption temperature of the volatile species. In these experiments the volatile species diffuses through the ice and subsequently desorbs; the diffusion is then traced by the decreasing infrared absorption. These experiments are usually not able to distinguish well between the mechanisms of diffusion as the molecule can diffuse from both weakly and strongly bound sites. In our method we monitor diffusion by observing changes in the CO$_2$ lattice IR modes. This method resembles that used by \citet{Lauck2015}, where the IR feature of the diffusing molecule, CO, was monitored, but presents a number of advantages.\\
\indent First, this technique could be extended to study ice systems in which the diffusing molecule itself is IR inactive, e.g. O$_2$, N$_2$, but produces a still produces a shift in the CO$_2$ LO mode upon mixing with CO$_2$. \\
\indent Second, LO phonon modes are very sensitive to the exact mixing morphology of the ice, which enabled us to distinguish between mixing through pore diffusion and mixing through bulk diffusion.\\
\indent Using LO phonons to trace ice mixing and diffusion also present some unique challenges. The LO phonon frequency shifts are the result of changes in the ice lattice to intermolecular forces between CO$_2$ and the diffusing CO molecules and there are potentially other processes that can also change the lattice structure. In particular, 
at temperatures similar to those employed in our experiments, the CO$_2$ ice may undergo pore collapse or reorganization, which could change the LO phonon frequency. To check the potential impact of CO$_2$ morphology changes, we also ran an isothermal experiment in which we deposited pure CO$_2$ ice. 
In this experiment we did not see a redshift of the LO mode indicating that CO diffusion into CO$_2$ is indeed responsible for the observed redshift during the diffusion experiments. At longer time scales, we did, however, see a slow blueshift and narrowing of the LO mode develop attributed to CO$_2$ crystallization. Fortunately, the CO$_2$ crystallization is slower than the CO diffusion process in our temperature regime and it appears that CO diffusion into CO$_2$ further slows the CO$_2$ crystallization rate. \\

\subsection{Diffusion and Desorption Barriers and their Astrophysical Implications}
The CO:CO$_2$ diffusion barriers extracted in this work, combined with the complementary measurements of the CO-CO$_2$ desorption barriers places CO:CO$_2$ ice diffusion into a growing family of systems with low, $<$0.3, ice diffusion-desorption barrier ratios. This suggests that diffusion may be underestimated in current gas-grain astrochemical models which typically adopt diffusion-desorption energy barrier ratios of 0.3 or higher \citep{Katz1999,Ruffle2000,Garrod2011,Chang2012}. However, it is important to note that these low diffusion barriers are only valid for ices with pores, and may be sensitive to porosity differences between laboratory and interstellar ices \citep{Garrod2013b}.\\
\indent It is further important to note that the diffusion-desorption barrier ratio for the CO:CO$_2$ system is larger by a factor of two compared to the diffusion-desorption barrier ratio for the CO:H$_2$O system that can be derived from experiments of \citet{Lauck2015} and \citet{Fayolle2016}. This strongly suggests 
that there is no universal ratio that can be applied in models, but rather that experiments and molecular dynamics models are needed for several other major ice constituents and for mixed ices to evaluate the range of possible ratios. \\
\indent Through our measurements to determine the diffusion-desorption barrier ratio for the CO:CO$_2$ ice system, we systematically measured CO-CO$_2$ desorption barriers for the first time. We found that the CO-CO$_2$ E$_{des}$ barriers are substantially higher than the previous estimates of \citet{Cleeves2014}, who report a CO-CO$_2$ desorption energy of 1110 K based on the peak desorption temperature of CO from CO$_2$. Considering only the individual pairwise interactions between CO and the ice substrate, we would expect that the CO desorption energy from water ice should be higher than that from CO$_2$. Instead, the ice morphology appears to be more important in controlling the CO desorption from H$_2$O and CO$_2$ ices. This implies that the CO desorption temperature in, for example, protoplanetary disks, may be high even when CO is not in direct contact with water ice. A recent study places the CO snowline in the iconic protoplanetary disk TW Hya at 22 AU \citep{Zhang2017} and explains its location as a result of CO binding directly to water ice. Our results show that the same desorption temperature could result from binding to CO$_2$ ice, which might be a likelier scenario when considering the freeze-out temperatures and chemistry of H$_2$O, CO$_2$ and CO.

\section{Conclusions}\label{sec:conclude}

In this work we report the diffusion of CO into CO$_2$ from initially layered ices at low temperatures. We make the following conclusions:
\begin{enumerate}
\item We show that the CO$_2$ $\nu_3$ LO phonon mode can be used to trace CO diffusion. This system could be used to study mixing phenomena between other astrophysically relevant ice constituents and CO$_2$.
\vspace{-3mm}
\item The diffusion coefficients depend on temperature as well as the CO:CO$_2$ ice thickness ratio.
\vspace{-3mm}
\item The temperature-dependent rates CO diffusion through CO$_2$ ice are well fit by an Arrhenius Law, which allows us to derive a diffusion barrier of 300 $\pm$ 40 K. 
\vspace{-3mm}
\item The CO from CO$_2$ desorption energies range from 1239-1407 K depending on the CO$_2$ ice deposition temperature. Some of the CO-CO$_2$ desorption barriers are similar to those from water ices, demonstrating that CO binds equally well to compact CO$_2$ as it does to compact water ice. 
\vspace{-3mm}
\item Combining these sets of experiments, we derive a diffusion-desorption barrier ratio for CO:CO$_2$ ices of 0.21--0.24. This ratio is low compared to what has been used in astrochemical models, suggesting that diffusion driven processes may be more efficient than what is currently assumed.
\vspace{-3mm}
\item The low diffusion barrier, combined with constraints on the diffusion kinetics supports a scenario where CO diffusion into CO$_2$ occurs along internal pores and across the CO$_2$ ice surface rather than through the bulk ice. The CO mobility and mixing in CO$_2$ ices depends on the number of surface binding sites resulting in a accumulation of CO at the CO$_2$ ice surface. 
\end{enumerate}

We gratefully acknowledge productive discussions with Rob Garrod, Eric Herbst, Aspen Clements, Matthew Reish and Shiliang Ma, as well as helpful comments from an anonymous reviewer. I.R.C. acknowledges support from the Sidney M. Hecht Graduate Fellowship. K.I.\"{O}. acknowledges funding from the Simons Collaboration on the Origins of Life (SCOL) and the David and Lucile Packard Foundation.

\end{document}